\documentclass[10pt]{article}
\usepackage{graphicx}
\usepackage{amsmath}
\usepackage{amssymb}
\usepackage{caption2}
\begin{document}
\begin{center}
{\large {\bf \sc{  Assignments  of the $X(4140)$, $X(4500)$, $X(4630)$ and  $X(4685)$ based on  the QCD sum rules  }}} \\[2mm]
Zhi-Gang  Wang \footnote{E-mail: zgwang@aliyun.com.  }   \\
 Department of Physics, North China Electric Power University, Baoding 071003, P. R. China
\end{center}

\begin{abstract}
In this article, we take into account our previous calculations based on the QCD sum rules, and tentatively  assign the $X(4630)$ as the $D_s^*\bar{D}_{s1}-D_{s1}\bar{D}_s^*$ tetraquark molecular state or   $[cs]_P[\bar{c}\bar{s}]_A+[cs]_A[\bar{c}\bar{s}]_P$ tetraquark state with the $J^{PC}=1^{-+}$, and assign the $X(3915)$ and $X(4500)$ as the 1S and 2S $[cs]_A[\bar{c}\bar{s}]_A$ tetraquark states respectively with the  $J^{PC}=0^{++}$.  Then we extend our  previous works to investigate  the LHCb's new tetraquark candidate $X(4685)$ as the first radial excited state of the $X(4140)$ with the QCD sum rules, and obtain the mass $M_{X}=4.70\pm0.12\,\rm{GeV}$, which is in very good agreement with  the experimental value $4684 \pm 7 {}^{+13}_{-16}\,\rm{MeV}$. Furthermore,   we investigate  the two-meson scattering state contributions  in details, and observe that the two-meson scattering states alone cannot saturate the QCD sum rules,  the contributions of the tetraquark states play an un-substitutable role, we can saturate the QCD sum rules with or without the two-meson scattering states.
\end{abstract}



\section{Introduction}

In 2009,  the CDF collaboration observed  an evidence for the $X(4140)$ in the $J/\psi\phi$  mass spectrum for the first time  with a significance of larger
 than $3.8 \sigma$  \cite{CDF0903}.  Subsequently, the existence of the $X(4140)$ was confirmed by the CDF, CMS and D0 collaborations  \cite{CDF1101,CMS1309,D0-1309,D0-1508}.
In 2016, the LHCb collaboration accomplished  the first full amplitude analysis of the $B^+\to J/\psi \phi K^+$ decays  and acquired a good description of the experimental data in the 6D phase space, and confirmed the $X(4140)$ and $X(4274)$  and determined the spin-parity-charge-conjugation   $J^{PC} =1^{++}$ \cite{LHCb-16061,LHCb-16062}. Furthermore,  the LHCb collaboration also  observed  two new exotic hadrons $X(4500)$ and $X(4700)$ in the $J/\psi \phi$ mass spectrum   and determined the    quantum numbers   $J^{PC} =0^{++}$ \cite{LHCb-16061,LHCb-16062}. The Breit-Wigner   masses and widths are
\begin{flalign}
 & X(4140) : M = 4146.5 \pm 4.5 ^{+4.6}_{-2.8} \mbox{ MeV}
\, , \, \Gamma = 83 \pm 21 ^{+21}_{-14} \mbox{ MeV} \, , \nonumber\\
 & X(4274) : M = 4273.3 \pm 8.3 ^{+17.2}_{-3.6} \mbox{ MeV}
\, , \, \Gamma = 56 \pm 11 ^{+8}_{-11} \mbox{ MeV} \, ,\nonumber \\
 & X(4500) : M = 4506 \pm 11 ^{+12}_{-15} \mbox{ MeV} \, ,
\, \Gamma = 92 \pm 21 ^{+21}_{-20} \mbox{ MeV} \, , \nonumber\\
 & X(4700) : M = 4704 \pm 10 ^{+14}_{-24} \mbox{ MeV} \, ,
\, \Gamma = 120 \pm 31 ^{+42}_{-33} \mbox{ MeV} \, .
\end{flalign}
Recently, the LHCb collaboration accomplished an improved full amplitude analysis of the $B^+\to J/\psi \phi K^+$ decays  using 6 times larger signal yields  than previously analyzed and observed a hidden-charm and hidden-strange tetraquark candidate  $X(4685)$  ($X(4630)$) in the mass spectrum of the $J/\psi \phi$ with a significance of $15\sigma$  ($5.5\sigma$),  the favored assignment of the spin-parity is $J^P=1^+$($1^-$),  the Breit-Wigner   mass and width are $4684 \pm 7 {}^{+13}_{-16}\,\rm{MeV}$  ($4626 \pm 16 {}^{+18}_{-110}\,\rm{MeV}$) and $126 \pm 15 {}^{+37}_{-41}\,\rm{MeV}$  ($174 \pm 27 {}^{+134}_{-73}\,\rm{MeV}$), respectively \cite{LHCb-X4685}. Furthermore, the
LHCb collaboration also observed two new tetraquark (molecular) state candidates $Z_{cs}(4000)$ and $Z_{cs}(4220)$ in the mass spectrum of the $J/\psi K^+$ with the preferred spin-parity $J^P=1^+$, and updated  the experimental values of the masses and widths of the $X(4500)$ and $X(4700)$ \cite{LHCb-X4685}.
  The $X(4140)$, $X(4274)$, $X(4500)$, $X(4630)$, $X(4685)$ and $X(4700)$ were observed in the mass spectrum of the  $J/\psi\phi$, their quantum numbers are   $J^{PC}=0^{++}$, $1^{++}$, $2^{++}$ for the S-wave couplings, and $0^{-+}$, $1^{-+}$, $2^{-+}$, $3^{-+}$ for the P-wave couplings.
In the present  work,  we discuss the possible assignments  of the $X(4140)$, $X(4500)$, $X(4630)$ and  $X(4685)$ based on  the QCD sum rules.

The article is arranged as follows: in Sect.2, we discuss the possible assignments of the $X(4630)$ and $X(4500)$ based on the QCD sum rules; in Sect.3, we get the QCD sum rules for the masses and pole residues of  the
 tetraquark  states $X(4140)/X(4685)$ with the $J^{PC}=1^{++}$;  in Sect.4, we obtain  numerical results and give discussions; and Sect.5 is aimed to get a
conclusion.

\section{Possible assignments of the $X(4630)$ and $X(4500)$ based on the QCD sum rules}
In this Section, we discuss the possible assignments of the $X(4630)$ and $X(4500)$ according to our previous calculations with the QCD sum rules.

In Ref.\cite{WZG-Landau}, we construct the color-singlet-color-singlet type four-quark current $J_{\mu\nu}(x)$ to investigate  the $D_s^*\bar{D}_{s1}-D_{s1}\bar{D}_s^*$ molecular state,
  \begin{eqnarray}
J_{\mu\nu}(x)&=&\frac{1}{\sqrt{2}}\Big[\bar{s}(x)\gamma_\mu c(x) \bar{c}(x)\gamma_\nu\gamma_5  s(x)-\bar{s}(x)\gamma_\nu\gamma_5 c(x) \bar{c}(x)\gamma_\mu s(x)\Big]\, . \end{eqnarray}
The current $J_{\mu\nu}(x)$ has definite charge conjugation $C=1$ but has not definite parity, the components $J_{0i}(x)$ and
  $J_{ij}(x)$ have positive-parity and negative-parity, respectively,  where the space  indexes $i$, $j=1$, $2$, $3$.
  The  neutral    current $J_{\mu\nu}(x)$ couples potentially to the $D_s^*\bar{D}_{s1}-D_{s1}\bar{D}_s^*$ two-meson scattering states or tetraquark molecular states $X_{D_s^*\bar{D}_{s1}-D_{s1}\bar{D}_s^*}$ with the quantum numbers $J^{PC}=1^{++}$ and $1^{-+}$, where we use the symbols  $D_s^*$ and $D_{s1}$ to represent the color-neutral clusters with the same quantum numbers as the physical  $D_s^*$ and $D_{s1}$ mesons, respectively. In the QCD sum rules, we choose the local currents, it is  better to call the $X_{D_s^*\bar{D}_{s1}-D_{s1}\bar{D}_s^*}$  as the color-singlet-color-singlet type  tetraquark state than call it as the tetraquark molecular state.
The traditional hidden-flavor mesons, such as the $q\bar{q}$, $c\bar{c}$ and $b\bar{b}$ quarkonia, have the normal quantum numbers $J^{PC}=0^{-+}$, $0^{++}$, $1^{--}$, $1^{+-}$, $1^{++}$, $2^{--}$, $2^{-+}$, $2^{++}$, $\cdots$.
The components  $J_{0i}(x)$ and  $J_{ij}(x)$ couple potentially to the $J^{PC}=1^{++}$ and $1^{-+}$ tetraquark molecular states, respectively. We construct projection   operators to project out  the contribution of the $J^{PC}=1^{-+}$ component unambiguously, and explore the $D_s^*\bar{D}_{s1}-D_{s1}\bar{D}_s^*$ tetraquark molecular state with the exotic quantum numbers  $J^{PC}=1^{-+}$ using the QCD sum rules, and acquire the prediction \cite{WZG-Landau},
\begin{eqnarray}
M_{X}&=&4.67\pm0.08\,\rm{GeV}\, ,
\end{eqnarray}
 which happens to coincide with the mass of the $X(4630)$ from the LHCb  collaboration, $M_{X(4630)}=4626 \pm 16 {}^{+18}_{-110}\,\rm{MeV}$ \cite{LHCb-X4685}.

The calculations based on the Bethe-Salpeter equation combined with the heavy meson effective Lagrangian also indicate that there exists such a  $D_s^*\bar{D}_{s1}-D_{s1}\bar{D}_s^*$ tetraquark molecular state with the exotic quantum numbers  $J^{PC}=1^{-+}$ \cite{FKGuo-Progr,DYChen-EPJC-2020}.
The predictions in Refs.\cite{WZG-Landau,FKGuo-Progr,DYChen-EPJC-2020} were achieved  before the discovery of the $X(4630)$. Whether or not the predictions of the QCD sum rules are reliable, the experimental data can reply.
After the discovery of the $X(4630)$ by the LHCb collaboration, Yang et al study the  charmonium-like molecules with hidden-strange via  the one-boson exchange mechanism, and assign the
$X(4630)$ to be the $D_s^*\bar{D}_{s1}$ molecular state with the quantum numbers $J^{PC}=1^{-+}$ \cite{LiuX-X4630}.

As long as the diquark-antidiquark type tetraquark states are concerned,  we usually take the  scalar ($S$), pseudoscalar ($P$), vector ($V$), axialvector ($A$)  and  tensor ($T$) diquark operators without introducing explicit P-waves as the  elementary  building blocks  to construct the interpolating currents. The tensor currents have both vector and axialvector components, and we construct projection  operators to project out the spin-parity $J^P=1^-$ and $1^+$ components explicitly, and denote  the corresponding operators as $\widetilde{V}$ and $\widetilde{A}$ respectively to avoid  ambiguity.
In Ref.\cite{WZG-ESF-EPJC-2874}, we choose the diquark-antidiquark type vector currents $J^{-}_\mu(x)$ and $J^{+}_\mu(x)$,
\begin{eqnarray}
J^{-}_\mu(x)&=&\frac{\varepsilon^{ijk}\varepsilon^{imn}}{\sqrt{2}}\left\{s^j(x)C c^k(x) \bar{s}^m(x)\gamma_\mu C \bar{c}^n(x)-s^j(x)C\gamma_\mu c^k(x)\bar{s}^m(x)C \bar{c}^n(x) \right\} \, , \nonumber\\
J^{+}_\mu(x)&=&\frac{\varepsilon^{ijk}\varepsilon^{imn}}{\sqrt{2}}\left\{s^j(x)C c^k(x) \bar{s}^m(x)\gamma_\mu C \bar{c}^n(x)+s^j(x)C\gamma_\mu c^k(x)\bar{s}^m(x)C \bar{c}^n(x) \right\} \, ,
\end{eqnarray}
to interpolate the $[cs]_P[\bar{c}\bar{s}]_A-[cs]_A[\bar{c}\bar{s}]_P$-type and $[cs]_P[\bar{c}\bar{s}]_A+[cs]_A[\bar{c}\bar{s}]_P$-type  tetraquark states  with the quantum numbers $J^{PC}=1^{--}$ and $1^{-+}$, respectively, and investigate their properties with the QCD sum rules, where the $i$, $j$, $k$, $m$, $n$ are color indexes. We acquire the predictions,
\begin{eqnarray}
M_{X(4630)}&=&4.63^{+0.11}_{-0.08}\, {\rm{GeV}\, \, \,\rm{for}}\,\,\, J^{PC}=1^{-+}\, ,\nonumber \\
M_{Y(4660)}&=&4.70^{+0.14}_{-0.10}\, {\rm{GeV}\, \, \, \rm{for}}\,\,\,J^{PC}=1^{--}\, ,
\end{eqnarray}
which happen to coincide with the masses of the $X(4630)$ and $Y(4660)$, respectively \cite{WZG-ESF-EPJC-2874}, and support assigning the $X(4630)$ and $Y(4660)$ to be the  tetraquark states with the symbolic quark constituents $c\bar{c}s\bar{s}$ and  with the quantum numbers $J^{PC}=1^{-+}$ and  $1^{--}$, respectively. The prediction of the mass $4.63^{+0.11}_{-0.08}\, {\rm{GeV}}$ was achieved  long before the discovery of the $Y(4630)$.

In Refs.\cite{X3915-X4500-EPJC-WZG, X3915-X4500-EPJA-WZG}, we construct the diquark-antidiquark type currents to explore the $[cs]_S[\bar{c}\bar{s}]_S$, $[cs]_P[\bar{c}\bar{s}]_P$, $[cs]_A[\bar{c}\bar{s}]_A$ and $[cs]_V[\bar{c}\bar{s}]_V$ tetraquark states with the quantum numbers $J^{PC}=0^{++}$  concordantly via the QCD sum rules, the numerical results support  assigning   the  $X(3915)$ and $X(4500)$ to be the ground state and first radial excited state of the $[cs]_A[\bar{c}\bar{s}]_A$ tetraquark states respectively with the quantum numbers $J^{PC}=0^{++}$, and assigning  the $X(4700)$ to be the ground state $[cs]_V[\bar{c}\bar{s}]_V$ tetraquark states with the quantum numbers $J^{PC}=0^{++}$. Furthermore, we also obtain the potability that assigning the   $X(3915)$ to be the ground state  $[cs]_S[\bar{c}\bar{s}]_S$ tetraquark state with the $J^{PC}=0^{++}$ \cite{X3915-X4500-EPJA-WZG}.
Our predictions,
\begin{eqnarray}
M_{X(4500)}&=&4.50^{+0.08}_{-0.09}\,\rm{GeV} \, ,\nonumber\\
M_{X(4700)}&=&4.70^{+0.08}_{-0.09}\,\rm{GeV} \, ,
\end{eqnarray}
are in very good  agreement with the LHCb improved measurement
$M_{X(4500)}=4474 \pm 3 \pm 3\,\rm{MeV}$ and $M_{X(4700)}=4694 \pm 4 {}^{+16}_{- 3}\,\rm{MeV}$ \cite{LHCb-X4685}.
Other assignments of the $X(4500)$ and $X(4700)$, such as the D-wave $cs\bar{c}\bar{s}$ tetraquark states with the $J^P=0^+$ are also possible \cite{Chen-Chen-Zhu-4500}, more theoretical and experimental works are still needed to obtain definite conclusion.

\begin{table}
\begin{center}
\begin{tabular}{|c|c|c|c|c|c|c|c|}\hline\hline
   $J^{PC}$                    & 1S                 & 2S                      & energy gaps   \\ \hline
   $1^{+-}$                    & $Z_c(3900)$        & $Z_c(4430)$             & 591\,\rm{MeV}        \\ \hline
    $0^{++}$                   & $X(3915)$          & $X(4500)$               & 588\,\rm{MeV}     \\ \hline
    $1^{+-}$                   & $Z_c(4020)$        & $Z_c(4600)$             & 576\,\rm{MeV}    \\ \hline
   $1^{++}$                    & $X(4140)$          & $X(4685)$               & 566\,\rm{MeV}     \\ \hline \hline
\end{tabular}
\end{center}
\caption{ The energy gaps between the ground states and first radial excited states of the hidden-charm tetraquark states with the possible assignments. }\label{1S2S}
\end{table}

In summary, according to the (possible) quantum numbers, decay modes and energy gaps, we can assign the $X(3915)$ and $X(4500)$ as the ground state and  first radial excited state of the hidden-charm  tetraquark states with the $J^{PC}=0^{++}$ \cite{X3915-X4500-EPJC-WZG,X4140-tetraquark-Lebed}, assign
the $Z_c(3900)$ and $Z_c(4430)$   as  the ground state and first radial excited state of the hidden-charm tetraquark states with the $J^{PC}=1^{+-}$, respectively \cite{Maiani-Z4430-1405,Nielsen-1401,WangZG-Z4430-CTP},   and assign the $Z_c(4020)$ and $Z_c(4600)$ as the ground state and  first radial excited  state of the hidden-charm tetraquark states with the $J^{PC}=1^{+-}$, respectively  \cite{ChenHX-Z4600-A,WangZG-axial-Z4600}. If we assign the $X(4685)$ to be the first radial excited state of the $X(4140)$ tentatively, we can get the energy gap $566\,\rm{MeV}$, it is reasonable, see Table \ref{1S2S}.

Moreover, in Ref.\cite{X4140-tetraquark-Lebed}, R. F. Lebed  and A. D. Polosa assign the $X(3915)$ and $X(4140)$ to  be the $J^{PC}=0^{++}$ and $1^{++}$ diquark-antidiquark type hidden-charm tetraquark states  $[cs]_{S}[\bar{c}\bar{s}]_{S}$ and $[cs]_{A}[\bar{c}\bar{s}]_{S}+[cs]_{S}[\bar{c}\bar{s}]_{A}$ respectively  based on the effective  Hamiltonian with the spin-spin and spin-orbit  interactions.
 In Ref.\cite{WZG-Di-X4140-EPJC}, we construct   the $[sc]_{\widetilde{A}}[\bar{s}\bar{c}]_A+[sc]_A[\bar{s}\bar{c}]_{\widetilde{A}}$ type  and $[sc]_{\widetilde{V}}[\bar{s}\bar{c}]_V-[sc]_V[\bar{s}\bar{c}]_{\widetilde{V}}$ type axialvector currents with the quantum numbers $J^{PC}=1^{++}$ to interpolate
   the $X(4140)$, and observe that only the $[sc]_{\widetilde{V}}[\bar{s}\bar{c}]_V-[sc]_V[\bar{s}\bar{c}]_{\widetilde{V}}$ type current can reproduce the mass and width of the $Y(4140)$ in a consistent way.

\section{The  $X(4140)/X(4685)$ as the $\rm{1S / 2S}$  axialvector tetraquark states}
In this Section, we extend our previous work \cite{WZG-Di-X4140-EPJC} to investigate  the $X(4685)$ as the first radial excitation of the $X(4140)$ with the QCD sum rules, and discuss the possible assignment of the $X(4685)$ as the tetraquark  state having  the quantum numbers $J^{PC}=1^{++}$.

Firstly, we write down  the two-point correlation function $\Pi_{\mu\mu^\prime}(p)$  in the QCD sum rules,
\begin{eqnarray}
\Pi_{\mu\mu^\prime}(p)&=&i\int d^4x e^{ip \cdot x} \langle0|T\left\{J_\mu(x)J_{\mu^\prime}^{\dagger}(0)\right\}|0\rangle \, ,
\end{eqnarray}
where
\begin{eqnarray}
J_\mu(x)&=&\frac{\varepsilon^{ijk}\varepsilon^{imn}}{\sqrt{2}}\left[s^{Tj}(x)C\sigma_{\mu\nu} c^k(x)\bar{s}^m(x)\gamma_5\gamma^\nu C \bar{c}^{Tn}(x)-s^{Tj}(x)C\gamma^\nu \gamma_5c^k(x)\bar{s}^m(x) \sigma_{\mu\nu} C \bar{c}^{Tn}(x) \right] \, . \nonumber\\
\end{eqnarray}
    The current $J_\mu(x)$ couples potentially to the $[sc]_{\widetilde{V}}[\bar{s}\bar{c}]_V-[sc]_V[\bar{s}\bar{c}]_{\widetilde{V}}$
  tetraquark states with the $J^{PC}=1^{++}$.
The tensor diquark  operator $\varepsilon^{ijk}s^{Tj}(x)C\sigma_{\mu\nu}c^k(x)$ has both the spin-parity $J^P=1^+$ and $1^-$ components, we project out the  $1^-$ component via  multiplying the tensor diquark operator by the vector antidiquark operator $\varepsilon^{imn}\bar{s}^m(x)\gamma_5\gamma^\nu C \bar{c}^{Tn}(x)$. In Ref.\cite{WZG-Di-X4140-EPJC}, we observe that the current $J_\mu(x)$ can reproduce the mass and width of the $Y(4140)$ satisfactorily.

At the hadron side,   we isolate the ground state ($X$) and first radial excited state ($X^\prime$) contributions, which are supposed to be the pole contributions from the $X(4140)$ and $X(4685)$,  respectively,
\begin{eqnarray}
\Pi_{\mu\mu^\prime}(p)&=&\left(\frac{\lambda_{X}^2}{M^2_{X}-p^2}+\frac{\lambda_{X^\prime}^2}{M^2_{X^\prime}-p^2}+\cdots\right)\left(-g_{\mu\mu^{\prime} } +\frac{p_\mu p_{\mu^{\prime}}}{p^2}\right) +\cdots  \nonumber\\
&=&\Pi(p^2)\left(-g_{\mu\mu^{\prime} } +\frac{p_\mu p_{\mu^{\prime}}}{p^2}\right) +\cdots \, ,
\end{eqnarray}
where the pole residues or decay constants  $\lambda_{X^{(\prime)}}$ are defined by $\langle 0|J_\mu(0)|X^{(\prime)}(p)\rangle=\lambda_{X^{(\prime)}}\, \varepsilon_\mu$,
the $\varepsilon_\mu$ are  the polarization vectors of the axialvector tetraquark states $X^{(\prime)}$.

A hadron, such as the usually called quark-antiquark type meson, tree-quark type baryon, diquark-antidiquark type tetraquark state, diquark-diquark-antiquark type pentaquark state, etc,  has definite quantum numbers and more than one Fock states. Any current operator  with the same quantum numbers and same quark structure as a Fock state in the hadron couples potentially to this  hadron, in other words, it has non-vanishing coupling to this hadron. Generally speaking,  we can construct several current operators to interpolate a hadron, or construct a current operator to interpolate several hadrons. Actually, a hadron has one or two main Fock states,
we call a hadron as a tetraquark  state if its main Fock component is of the diquark-antidiquark type.

In the present work, the diquark-antidiquark type local four-quark current operator $J_\mu(x)$ having  the  quantum numbers $J^{PC}=1^{++}$ couples potentially  to the
diquark-antidiquark type tetraquark states with the same quantum numbers $J^{PC}=1^{++}$. On the other hand,  this  local current $J_\mu(x)$  can be re-arranged into a special superposition of  a series of color-singlet-color-singlet type  currents through the Fierz transformation both in the Dirac spinor space and color space,
\begin{eqnarray}\label{Fierz}
2\sqrt{2}J_\mu(x) &=&-\bar{c}(x)\sigma_{\mu\nu}\gamma_5 c(x)\,\bar{s}(x)\gamma^\nu s(x)+\bar{s}(x)\sigma_{\mu\nu}\gamma_5 s(x)\,\bar{c}(x)\gamma^\nu c(x)-3i\bar{c}(x)\gamma_\mu\gamma_5c(x)\,\bar{s}(x)s(x)\nonumber\\
&&+3i\bar{s}(x)\gamma_\mu\gamma_5s(x)\,\bar{c}(x)c(x)
-\bar{c}(x)\sigma_{\mu\nu} s(x)\,\bar{s}(x)\gamma^\nu\gamma_5 c(x)+\bar{s}(x)\sigma_{\mu\nu} c(x)\,\bar{c}(x)\gamma^\nu\gamma_5 s(x)\nonumber\\
&&-3\bar{c}(x)\gamma_\mu s(x)\,\bar{s}(x)i\gamma_5c(x)
+3\bar{s}(x)\gamma_\mu c(x)\,\bar{c}(x)i\gamma_5s(x)\, ,
\end{eqnarray}
which couple potentially  to the tetraquark molecular states or  two-meson scattering  states  having the quantum numbers  $J^{PC}=1^{++}$.
The diquark-antidiquark type tetraquark states can be viewed  as a special superposition of  a series of color-singlet-color-singlet molecular states and embody the net effects, and vise versa.

The diquark-antidiquark type tetraquark state can  be plausibly described by two diquarks in a double well potential
which are separated by a barrier \cite{Wilczek-diquark,Polosa-diquark}, the spatial distance between the diquark and antidiquark leads  to smaller  wave-function overlap between the quark and antiquark constituents,  the repulsive barrier or spatial distance  frustrates the Fierz rearrangements or recombinations between  the quarks and antiquarks, therefore suppresses  hadronizing   to the meson-meson pairs  \cite{Wilczek-diquark,Polosa-diquark,Brodsky-PRL,WZG-IJMPA-nonlocal}.

If the color-singlet-color-singlet type components in Eq.\eqref{Fierz}, such as  $\bar{c}(x)\sigma_{\mu\nu}\gamma_5 c(x)\,\bar{s}(x)\gamma^\nu s(x)$, $\bar{s}(x)\sigma_{\mu\nu}\gamma_5 s(x)\,\bar{c}(x)\gamma^\nu c(x)$, etc, only couple potentially  to the two-meson (TM) scattering states,
we obtain the correlation function $\Pi_{TM}(p^2)$ at the hadron side,
\begin{eqnarray}
\Pi_{TM}(p^2) &=&\frac{1}{768\pi^2} \int_{m_{J/\psi\phi}^2}^{\infty} ds\frac{1}{s-p^2}\frac{\lambda^{\frac{1}{2}}\left(s,m^2_{J/\psi},m^2_{\phi}\right)}{s} \overline{\rho}_{J/\psi\phi}(s) \nonumber\\
&&+\frac{1}{1536\pi^2}f^2_{\phi}m^2_{\phi} f^2_{h_c} \int_{m_{h_c\phi}^2}^{\infty} ds\frac{1}{s-p^2}\frac{\lambda^{\frac{1}{2}}\left(s,m^2_{h_c},m^2_{\phi}\right)}{s} \overline{\rho}_{h_c\phi}(s) \nonumber\\
&&+\frac{1}{1536\pi^2}f^2_{J/\psi}m^2_{J/\psi} f^2_{h_s} \int_{m_{h_sJ/\psi}^2}^{\infty} ds\frac{1}{s-p^2}\frac{\lambda^{\frac{1}{2}}\left(s,m^2_{h_s},m^2_{J/\psi}\right)}{s} \overline{\rho}_{h_sJ/\psi}(s) \nonumber\\
&&+\frac{3}{512\pi^2}f^2_{f_0}m^2_{f_0} f^2_{\chi_{c1}}m^2_{\chi_{c1}} \int_{m_{f_0\chi_{c1}}^2}^{\infty} ds\frac{1}{s-p^2}\frac{\lambda^{\frac{1}{2}}\left(s,m^2_{\chi_{c1}},m^2_{f_0}\right)}{s} \overline{\rho}_{\chi_{c1}f_0}(s) \nonumber\\
&&+\frac{3}{512\pi^2}f^2_{f_1}m^2_{f_1} f^2_{\chi_{c0}}m^2_{\chi_{c0}} \int_{m_{f_1\chi_{c0}}^2}^{\infty} ds\frac{1}{s-p^2}\frac{\lambda^{\frac{1}{2}}\left(s,m^2_{\chi_{c0}},m^2_{f_1}\right)}{s} \overline{\rho}_{\chi_{c0}f_1}(s) \nonumber\\
&&+\frac{3}{512\pi^2}f^2_{f_0}m^2_{f_0} f^2_{\eta_{c}} \int_{m_{f_0\eta_{c}}^2}^{\infty} ds\frac{1}{s-p^2}\frac{\lambda^{\frac{3}{2}}\left(s,m^2_{\eta_{c}},m^2_{f_0}\right)}{s^2} \nonumber\\
&&+\frac{3}{512\pi^2}f^2_{\chi_{c0}}m^2_{\chi_{c0}} f^2_{\eta} \int_{m_{\chi_{c0}\eta}^2}^{\infty} ds\frac{1}{s-p^2}\frac{\lambda^{\frac{3}{2}}\left(s,m^2_{\chi_{c0}},m^2_{\eta}\right)}{s^2}   \nonumber\\
&&+\frac{1}{768\pi^2}f^2_{D_{s1}}m^2_{D_{s1}} f^{2}_{T,D_{s}^*} \int_{m^2_{D_{s1}D^*_s}}^{\infty} ds\frac{1}{s-p^2}\frac{\lambda^{\frac{1}{2}}\left(s,m^2_{D_{s1}},m^2_{D_s^*}\right)}{s} \overline{\rho}_{D_{s1}D^*_s}(s) \nonumber\\
&&+\frac{1}{192\pi^2} \int_{m^2_{D_{s}D^*_s}}^{\infty} ds\frac{1}{s-p^2}\frac{\lambda^{\frac{1}{2}}\left(s,m^2_{D_{s}},m^2_{D_s^*}\right)}{s} \overline{\rho}_{D_{s}D^*_s}(s) \nonumber
\end{eqnarray}
\begin{eqnarray}
&&+\frac{3}{256\pi^2} \frac{f_{D_s}^2m_{D_s}^4f_{D_{s0}}^2}{m_c^2}\int_{m^2_{D_{s}D_{s0}}}^{\infty} ds\frac{1}{s-p^2} \frac{\lambda^{\frac{3}{2}}\left(s,m^2_{D_{s}},m^2_{D_{s0}}\right)}{s^2} \nonumber\\
&&+(J/\psi \phi \to \psi^\prime\phi)+(J/\psi\phi \to \psi^{\prime\prime}\phi)+(h_c\phi \to h_c^\prime\phi)+\cdots \, ,
\end{eqnarray}
where
\begin{eqnarray}
\overline{\rho}_{J/\psi\phi}(s)&=&f_\phi^2m_\phi^2f_{T,J/\psi}^2\left(-s+8{m^2_{J/\psi}}                                    -{m^2_\phi}+\frac{({m^2_{J/\psi}}-{m^2_\phi})^2}{s}+\frac{(s-{m^2_{J/\psi}})^2}{{m^2_\phi}} \right) \nonumber\\
&&+2f_{J/\psi}m_{J/\psi}{f_\phi}{m_\phi}{f_{T,\phi}}{f_{T,J/\psi}}\left(5s-4m^2_{J/\psi}-4{m^2_\phi}
-\frac{({m^2_{J/\psi}}-{m^2_\phi})^2}{s}\right)  \nonumber\\
&&+f_{J/\psi}^2m_{J/\psi}^2f^2_{T,\phi} \left(-s+8{m^2_\phi}-{m^2_{J/\psi}}+\frac{({m^2_{J/\psi}}-{m^2_\phi})^2}{s}+\frac{(s-{m^2_\phi})^2}{{m^2_{J/\psi}}} \right) \, ,
\end{eqnarray}
\begin{eqnarray}
\overline{\rho}_{h_c\phi}(s)&=&-2s-10{m^2_{h_c}}-2{m^2_\phi}
+\frac{2{m^4_{\phi}}-3{m^2_{h_c}}{m^2_{\phi}}}{s}+\frac{2s^2-3s{m^2_{h_c}}}{{m^2_\phi}}
+\frac{m^6_{h_c}}{s{m^2_\phi}} \, ,
\end{eqnarray}
\begin{eqnarray}
\overline{\rho}_{h_sJ/\psi}(s)&=&-2s-10{m^2_{h_s}}-2{m^2_{J/\psi}}
+\frac{2{m^4_{J/\psi}}-3{m^2_{h_s}}{m^2_{J/\psi}}}{s}+\frac{2s^2-3s{m^2_{h_s}}}{{m^2_{J/\psi}}}
+\frac{m^6_{h_s}}{s{m^2_{J/\psi}}} \, ,
\end{eqnarray}
\begin{eqnarray}
\overline{\rho}_{\chi_{c1}f_0}(s)&=&10+\frac{m^2_{\chi_{c1}}-2m^2_{f_0}}{s}
+\frac{s-2m^2_{f_0}}{m^2_{\chi_{c1}}}+\frac{m^4_{f_0}}{sm^2_{\chi_{c1}}}\, ,
\end{eqnarray}
\begin{eqnarray}
\overline{\rho}_{\chi_{c0}f_1}(s)&=&10+\frac{m^2_{f_1}-2m^2_{\chi_{c0}}}{s}
+\frac{s-2m^2_{\chi_{c0}}}{m^2_{f_1}}+\frac{m^4_{\chi_{c0}}}{sm^2_{f_1}}\, ,
\end{eqnarray}
\begin{eqnarray}
\overline{\rho}_{D_{s1}D^{*}_{s}}(s)&=&-2s-10m^2_{D^{*}_{s}}-2m^2_{D_{s1}}
+\frac{2m^4_{D_{s1}}-3m^2_{D^{*}_{s}}m^2_{D_{s1}}}{s}
+\frac{2s^2-3sm^2_{D^{*}_{s}}}{m^2_{D_{s1}}}+\frac{m^6_{D^{*}_{s}}}{sm^2_{D_{s1}}}\, ,
\end{eqnarray}
\begin{eqnarray}
\overline{\rho}_{D_{s}D^{*}_{s}}(s)&=&\left(\frac{f_{D_{s}}f_{T,D^{*}_{s}}(s-m^2_{D_{s}}-m^2_{D^{*}_{s}})
}{4}-\frac{3f_{D_{s}}m^2_{D_{s}}f_{D^{*}_{s}}m_{D^{*}_{s}}}{2m_c}\right)^2     \nonumber\\
&&\left(10+\frac{m^2_{D^{*}_{s}}-2m^2_{D_{s}}}{s}+\frac{s-2m^2_{D_{s}}}{m^2_{D^{*}_{s}}}+\frac{m^4_{D_{s}}}{sm^2_{D^{*}_{s}}}  \right)  \nonumber\\
&&-\left(\frac{f^2_{D_{s}}f^2_{T,D^{*}_{s}}(s-m^2_{D_{s}}-m^2_{D^{*}_{s}})
}{8}-\frac{3f^2_{D_{s}}m^2_{D_{s}}f_{D^{*}_{s}}m_{D^{*}_{s}}f_{T,D^{*}_{s}}}{4m_c}\right)\nonumber\\
&&\left(s+m^2_{D^{*}_{s}}+2m^2_{D_{s}}+\frac{3m^2_{D^{*}_{s}}m^2_{D_{s}}-m^4_{D^{*}_{s}}-3m^4_{D_{s}}}{s}
+\frac{3sm^2_{D_{s}}-s^2-3m^4_{D_{s}}}{m^2_{D^{*}_{s}}}+\frac{m^6_{D_{s}}}{sm^2_{D^{*}_{s}}}\right)\nonumber\\
&&+\frac{f^2_{D_{s}}f^2_{T,D^{*}_{s}}}{4}\left(-s^2-m^4_{D^{*}_{s}}+m^4_{D_{s}}+m^2_{D^{*}_{s}}m^2_{D_{s}}
+sm^2_{D_{s}}+\frac{3sm^2_{D^{*}_{s}}}{2}+\frac{m^8_{D_{s}}}{4sm^2_{D^{*}_{s}}} \right.\nonumber\\
&&\left.+\frac{6m^2_{D^{*}_{s}}m^4_{D_{s}}-4m^4_{D^{*}_{s}}m^2_{D_{s}}+m^6_{D^{*}_{s}}-4m^6_{D_{s}}}{4s}
+\frac{s^3-4s^2m^2_{D_{s}}+6sm^4_{D_{s}}-4m^6_{D_{s}}}{4m^2_{D^{*}_{s}}} \right)\, ,
\end{eqnarray}
$\lambda(a,b,c)=a^2+b^2+c^2-2ab-2ac-2bc$, $m_{J/\psi\phi}=m_{J/\psi}+m_{\phi}$, $m_{h_c\phi}=m_{h_c}+m_{\phi}$,
$m_{h_sJ/\psi}=m_{h_s}+m_{J/\psi}$, $m_{f_0\chi_{c1}}=m_{f_0}+m_{\chi_{c1}}$,
$m_{f_1\chi_{c0}}=m_{f_1}+m_{\chi_{c0}}$, $m_{f_0\eta_{c}}=m_{f_0}+m_{\eta_c}$,
$m_{\chi_{c0}\eta}=m_{\chi_{c0}}+m_{\eta}$,  $m_{D_{s1}D^*_s}=m_{D_{s1}}+m_{D_s^*}$, $m_{D_{s}D^*_s}=m_{D_{s}}+m_{D_s^*}$,
 $m_{D_{s}D_{s0}}=m_{D_s}+m_{D_{s0}}$,
 and we have taken the standard definitions of the decay constants,
\begin{eqnarray}
\langle 0|\bar{c}(0)\sigma_{\mu\nu}c(0)|J/\psi(p)\rangle&=&if^T_{J/\psi}\left(\varepsilon_\mu p_\nu-\varepsilon_{\nu}p_\mu \right)\, , \nonumber\\
\langle 0|\bar{s}(0)\sigma_{\mu\nu}s(0)|\phi(p)\rangle&=&if^T_{\phi}\left(\varepsilon_\mu p_\nu-\varepsilon_{\nu}p_\mu \right)\, , \nonumber\\
\langle 0|\bar{c}(0)\sigma_{\mu\nu}s(0)|D_s^*(p)\rangle&=&if^T_{D_s^*}\left(\varepsilon_\mu p_\nu-\varepsilon_{\nu}p_\mu \right)\, ,
\end{eqnarray}
\begin{eqnarray}
\langle 0|\bar{c}(0)\sigma_{\mu\nu}c(0)|h_c(p)\rangle&=&if_{h_c}\varepsilon_{\mu\nu\alpha\beta}\varepsilon^\alpha p^\beta\, , \nonumber\\
\langle 0|\bar{s}(0)\sigma_{\mu\nu}s(0)|h_s(p)\rangle&=&if_{h_s}\varepsilon_{\mu\nu\alpha\beta}\varepsilon^\alpha p^\beta\, ,
\end{eqnarray}
\begin{eqnarray}
\langle 0|\bar{c}(0)\gamma_{\mu}c(0)|J/\psi(p)\rangle&=&f_{J/\psi}m_{J/\psi}\varepsilon_\mu \, , \nonumber\\
\langle 0|\bar{s}(0)\gamma_{\mu}s(0)|\phi(p)\rangle&=&f_{\phi}m_{\phi}\varepsilon_\mu \, , \nonumber\\
\langle 0|\bar{c}(0)\gamma_{\mu}s(0)|D_s^*(p)\rangle&=&f_{D_s^*}m_{D_s^*}\varepsilon_\mu \, ,
\end{eqnarray}
\begin{eqnarray}
\langle 0|\bar{c}(0)\gamma_{\mu}\gamma_5c(0)|\chi_{c1}(p)\rangle&=&f_{\chi_{c1}}m_{\chi_{c1}}\varepsilon_\mu \, , \nonumber\\
\langle 0|\bar{s}(0)\gamma_{\mu}\gamma_5s(0)|f_1(p)\rangle&=&f_{f_1}m_{f_1}\varepsilon_\mu \, ,\nonumber\\
\langle 0|\bar{c}(0)\gamma_{\mu}\gamma_5s(0)|D_{s1}(p)\rangle&=&f_{D_{s1}}m_{D_{s1}}\varepsilon_\mu \, ,
\end{eqnarray}
\begin{eqnarray}
\langle 0|\bar{c}(0)\gamma_{\mu}\gamma_5c(0)|\eta_{c}(p)\rangle&=&if_{\eta_c}p_\mu \, , \nonumber\\
\langle 0|\bar{s}(0)\gamma_{\mu}\gamma_5s(0)|\eta(p)\rangle&=&if_{\eta}p_\mu \, , \nonumber\\
\langle 0|\bar{c}(0)\gamma_{\mu}\gamma_5s(0)|D_s(p)\rangle&=&if_{D_s}p_\mu \, ,
\end{eqnarray}
\begin{eqnarray}
\langle 0|\bar{c}(0)c(0)|\chi_{c0}(p)\rangle&=&f_{\chi_{c0}}m_{\chi_{c0}}\, , \nonumber\\
\langle 0|\bar{s}(0)s(0)|f_0(p)\rangle&=&f_{f_0}m_{f_0} \, ,
\end{eqnarray}
\begin{eqnarray}
\langle 0|\bar{c}(0)i\gamma_5s(0)|D_{s}(p)\rangle&=&\frac{f_{D_s}m^2_{D_s}}{m_c}\, ,
\end{eqnarray}
\begin{eqnarray}
\langle 0|\bar{c}(0)\gamma_{\mu}s(0)|D_{s0}(p)\rangle&=&f_{D_{s0}}p_\mu \, ,
\end{eqnarray}
the $\varepsilon_\mu$ are the polarization vectors of the vector and axialvector mesons.

We accomplish the  operator product expansion for the correlation function $\Pi_{\mu\mu^\prime}(p)$ up to the vacuum condensates of dimension 10 consistently \cite{WZG-HT-PRD-2014,WZG-HT-EPJC-2891,WZG-EPJC-2963}.
In calculations, we assume dominance of the  intermediate vacuum  state tacitly, just like in previous works \cite{WZG-HT-PRD-2014,WZG-HT-EPJC-2891,WZG-EPJC-2963}, and insert  the  intermediate vacuum
state alone  in all the channels,   vacuum saturation works well in the large $N_c$ limit \cite{Novikov--shifman}.
Up to now,  almost in all the QCD sum rules for the multiquark states,  vacuum saturation is assumed for the higher dimensional vacuum condensates, except  in some cases  the parameter $\varrho>1$, \begin{eqnarray}
 \langle0|:\bar{q}_{\alpha}^{i} q_{\beta}^{j}\bar{q}_{\lambda}^{m} q_{\tau}^{n}:|0\rangle
  &=&\frac{\varrho}{16N_c^2}\langle\bar{q} q\rangle^2
 \left(\delta_{ij}\delta_{mn}\delta_{\alpha\beta}\delta_{\lambda\tau}-\delta_{in}\delta_{jm}\delta_{\alpha\tau}\delta_{\beta\lambda} \right)\, ,
  \end{eqnarray}
 which parameterizes  deviations from the factorization hypothesis,  is introduced by hand
for the sake of fine-tuning  \cite{Narison-DvDv-D-8},  where $q=u$, $d$, $s$, the $i$, $j$, $m$ and $n$ are color indexes, the $\alpha$, $\beta$, $\lambda$ and $\tau$ are Dirac spinor indexes.

In the original works, Shifman,  Vainshtein and  Zakharov took
 the factorization hypothesis based on  two reasons \cite{SVZ79}.  The first one is the rather large value of the quark condensate $\langle\bar{q}q\rangle$,
 the second one is the duality between the quark and physical states, they reproduce each other,  counting both the quark and
physical states (beyond the vacuum states) maybe lead to  a double counting  \cite{SVZ79}.

In the QCD sum rules for the $q\bar{q}$, $q\bar{Q}$, $Q\bar{Q}$  mesons,  the $\langle\bar{q}q\rangle^2$  are always companied with the fine-structure constant $\alpha_s=\frac{g_s^2}{4\pi}$, and play a minor important (or tiny) role,
the deviation from $\varrho=1$, for example, $\varrho=2\sim 3$, cannot make much difference in the numerical predictions,
though in some cases the values  $\varrho>1$ can lead to better QCD sum rules  \cite{Review-rho-kappa,Narison-rho}.
 However, in the QCD sum rules for the multiquark states, the $\langle\bar{q}q\rangle^2$ play  an important role,
 large values, for example, if we take the value $\varrho=2$ in the present case, we can obtain the uncertainties $\delta M_{X}=+0.08\,\rm{GeV}$ and $\delta M_{X^\prime}=+0.09\,\rm{GeV}$, which are of the same order of the total uncertainties from other parameters.  Sometimes  large values of the $\varrho $ can destroy the platforms  in the QCD sum rules for the multiquark states \cite{WZG-DvDvDv}.

 The true values of the higher dimensional vacuum condensates remain unknown or poorly known, if the true values  $\varrho >1$ or $\gg 1$,  the QCD sum rules for the multiquark  states have considerably large systematic uncertainties and are
less reliable than those of the conventional  mesons and baryons \cite{Gubler-SB}. We just make predictions for the multiquark masses with the QCD sum rules based on  vacuum saturation, then confront them to the experimental data in the future to examine  the theoretical calculations.

After the analytical expression of the QCD spectral density was acquired, we take the quark-hadron duality below the continuum thresholds $s_0$ and $s_0^\prime$ by including the contributions of the 1S state and 1S plus 2S states, respectively,  and perform Borel transform  in regard to
the variable $P^2=-p^2$ to obtain  the two QCD sum rules:
\begin{eqnarray}\label{QCDSR-1}
\lambda^2_{X}\, \exp\left(-\frac{M^2_{X}}{T^2}\right)&=& \int_{4m_c^2}^{s_0} ds\, \rho(s) \, \exp\left(-\frac{s}{T^2}\right) \, , \nonumber\\
&=&\Pi(\tau)\, ,
\end{eqnarray}
\begin{eqnarray}\label{QCDSR-2}
\lambda^2_{X}\, \exp\left(-\frac{M^2_{X}}{T^2}\right)+\lambda^2_{X^\prime}\, \exp\left(-\frac{M^2_{X^\prime}}{T^2}\right)&=& \int_{4m_c^2}^{s_0^\prime} ds\, \rho(s) \, \exp\left(-\frac{s}{T^2}\right) \, ,\nonumber\\
&=&\Pi^{\prime}(\tau)\, ,
\end{eqnarray}
where $\tau=\frac{1}{T^2}$. For the explicit expression of the  spectral density  $\rho(s)$ at the quark and gluon level, one can consult Ref.\cite{WZG-Di-X4140-EPJC}.
 We define $D^n=\left( -\frac{d}{d\tau}\right)^n$ with $n=0$, $1$, $2$, $\cdots$, then   acquire the QCD sum rules for the masses,
 \begin{eqnarray}\label{QCDSR-1-M1}
M_{X}^2&=&\frac{D \Pi(\tau)}{\Pi(\tau)}\, ,
\end{eqnarray}
and
\begin{eqnarray}\label{QCDSR-2-M2}
M_{X^\prime}^2&=&\frac{b+\sqrt{b^2-4c} }{2} \, , \\
\lambda_{X^\prime}^2&=&\frac{\left(D-M_X^2\right)\Pi_{QCD}(\tau)}{M_{X^\prime}^2-M_X^2}\exp\left(\tau M_{X^\prime}^2 \right)\, , \nonumber
\end{eqnarray}
where
\begin{eqnarray}
b&=&\frac{D^3\otimes D^0-D^2\otimes D}{D^2\otimes D^0-D\otimes D}\, , \nonumber\\
c&=&\frac{D^3\otimes D-D^2\otimes D^2}{D^2\otimes D^0-D\otimes D}\, , \nonumber\\
D^j \otimes D^k&=&D^j\Pi^{\prime}(\tau) \,  D^k\Pi^{\prime}(\tau)\, ,
\end{eqnarray}
the indexes $i=1,2$ and $j,k=0,1,2,3$. For the technical details in  obtaining the QCD sum rules in Eq.\eqref{QCDSR-2-M2}, one can consult Refs.\cite{X3915-X4500-EPJC-WZG,WangZG-Z4430-CTP,Baxi-G}.

On the other hand, if we saturate the hadron side of the QCD sum rules with the contributions of the two-meson scattering sates, we obtain the following two QCD sum rules,
\begin{eqnarray}\label{TM-QCDSR-1}
\Pi_{TM}(T^2) &=&\frac{1}{768\pi^2} \int_{m_{J/\psi\phi}^2}^{s_0^\prime} ds\frac{\lambda^{\frac{1}{2}}\left(s,m^2_{J/\psi},m^2_{\phi}\right)}{s} \overline{\rho}_{J/\psi\phi}(s) \exp\left(-\frac{s}{T^2}\right)\nonumber\\
&&+\frac{1}{1536\pi^2}f^2_{\phi}m^2_{\phi} f^2_{h_c} \int_{m_{h_c\phi}^2}^{s_0^\prime} ds\frac{\lambda^{\frac{1}{2}}\left(s,m^2_{h_c},m^2_{\phi}\right)}{s} \overline{\rho}_{h_c\phi}(s)\exp\left(-\frac{s}{T^2}\right) \nonumber\\
&&+\frac{1}{1536\pi^2}f^2_{J/\psi}m^2_{J/\psi} f^2_{h_s} \int_{m_{h_sJ/\psi}^2}^{s_0^\prime} ds\frac{\lambda^{\frac{1}{2}}\left(s,m^2_{h_s},m^2_{J/\psi}\right)}{s} \overline{\rho}_{h_sJ/\psi}(s) \exp\left(-\frac{s}{T^2}\right)\nonumber\\
&&+\frac{3}{512\pi^2}f^2_{f_0}m^2_{f_0} f^2_{\chi_{c1}}m^2_{\chi_{c1}} \int_{m_{f_0\chi_{c1}}^2}^{s_0^\prime} ds\frac{\lambda^{\frac{1}{2}}\left(s,m^2_{\chi_{c1}},m^2_{f_0}\right)}{s} \overline{\rho}_{\chi_{c1}f_0}(s)\exp\left(-\frac{s}{T^2}\right) \nonumber\\
&&+\frac{3}{512\pi^2}f^2_{f_1}m^2_{f_1} f^2_{\chi_{c0}}m^2_{\chi_{c0}} \int_{m_{f_1\chi_{c0}}^2}^{s_0^\prime} ds\frac{\lambda^{\frac{1}{2}}\left(s,m^2_{\chi_{c0}},m^2_{f_1}\right)}{s} \overline{\rho}_{\chi_{c0}f_1}(s) \exp\left(-\frac{s}{T^2}\right)\nonumber\\
&&+\frac{3}{512\pi^2}f^2_{f_0}m^2_{f_0} f^2_{\eta_{c}} \int_{m_{f_0\eta_{c}}^2}^{s_0^\prime} ds\frac{\lambda^{\frac{3}{2}}\left(s,m^2_{\eta_{c}},m^2_{f_0}\right)}{s^2} \exp\left(-\frac{s}{T^2}\right)\nonumber\\
&&+\frac{3}{512\pi^2}f^2_{\chi_{c0}}m^2_{\chi_{c0}} f^2_{\eta} \int_{m_{\chi_{c0}\eta}^2}^{s_0^\prime} ds\frac{\lambda^{\frac{3}{2}}\left(s,m^2_{\chi_{c0}},m^2_{\eta}\right)}{s^2}\exp\left(-\frac{s}{T^2}\right)   \nonumber\\
&&+\frac{1}{768\pi^2}f^2_{D_{s1}}m^2_{D_{s1}} f^{2}_{T,D_{s}^*} \int_{m^2_{D_{s1}D^*_s}}^{s_0^\prime} ds\frac{\lambda^{\frac{1}{2}}\left(s,m^2_{D_{s1}},m^2_{D_s^*}\right)}{s} \overline{\rho}_{D_{s1}D^*_s}(s)\exp\left(-\frac{s}{T^2}\right) \nonumber\\
&&+\frac{1}{192\pi^2} \int_{m^2_{D_{s}D^*_s}}^{s_0^\prime} ds\frac{\lambda^{\frac{1}{2}}\left(s,m^2_{D_{s}},m^2_{D_s^*}\right)}{s} \overline{\rho}_{D_{s}D^*_s}(s) \exp\left(-\frac{s}{T^2}\right)\nonumber\\
&&+\frac{3}{256\pi^2} \frac{f_{D_s}^2m_{D_s}^4f_{D_{s0}}^2}{m_c^2}\int_{m^2_{D_{s}D_{s0}}}^{s_0^\prime} ds \frac{\lambda^{\frac{3}{2}}\left(s,m^2_{D_{s}},m^2_{D_{s0}}\right)}{s^2} \exp\left(-\frac{s}{T^2}\right)\nonumber\\
&&+(J/\psi \phi \to \psi^\prime\phi)+(J/\psi\phi \to \psi^{\prime\prime}\phi)+(h_c\phi \to h_c^\prime\phi)+\cdots  \nonumber\\
&=&\kappa\Pi^\prime(T^2)\, ,
\end{eqnarray}
\begin{eqnarray}\label{TM-QCDSR-2}
\frac{d}{d(1/T^2)}\Pi_{TM}(T^2) &=&\kappa \frac{d}{d(1/T^2)}\Pi^\prime(T^2)\, .
\end{eqnarray}
In Eqs.\eqref{TM-QCDSR-1}-\eqref{TM-QCDSR-2},  we introduce the parameter $\kappa$ to measure the deviations from $1$, if $\kappa\approx1$, we can acquire  the conclusion tentatively that the two-meson scattering states can  saturate the QCD sum rules.

\section{Numerical results and discussions}
At the QCD side, we choose  the conventional values of the vacuum condensates
$\langle\bar{q}q \rangle=-(0.24\pm 0.01\, \rm{GeV})^3$,  $\langle\bar{s}s \rangle=(0.8\pm0.1)\langle\bar{q}q \rangle$,
 $\langle\bar{s}g_s\sigma G s \rangle=m_0^2\langle \bar{s}s \rangle$,
$m_0^2=(0.8 \pm 0.1)\,\rm{GeV}^2$, $\langle \frac{\alpha_s
GG}{\pi}\rangle=(0.33\,\rm{GeV})^4 $    at the energy scale  $\mu=1\, \rm{GeV}$
\cite{SVZ79,Reinders85,ColangeloReview}, and  prefer the modified minimal subtracted  masses $m_{c}(m_c)=(1.275\pm0.025)\,\rm{GeV}$ and $m_s(\mu=2\,\rm{GeV})=(0.095\pm0.005)\,\rm{GeV}$
 from the Particle Data Group \cite{PDG}.
In addition,  we consider  the energy-scale dependence of the input parameters,
 \begin{eqnarray}
 \langle\bar{s}s \rangle(\mu)&=&\langle\bar{s}s \rangle({\rm 1 GeV})\left[\frac{\alpha_{s}({\rm 1 GeV})}{\alpha_{s}(\mu)}\right]^{\frac{12}{33-2n_f}}\, , \nonumber\\
   \langle\bar{s}g_s \sigma Gs \rangle(\mu)&=&\langle\bar{s}g_s \sigma Gs \rangle({\rm 1 GeV})\left[\frac{\alpha_{s}({\rm 1 GeV})}{\alpha_{s}(\mu)}\right]^{\frac{2}{33-2n_f}}\, ,\nonumber\\
m_c(\mu)&=&m_c(m_c)\left[\frac{\alpha_{s}(\mu)}{\alpha_{s}(m_c)}\right]^{\frac{12}{33-2n_f}} \, ,\nonumber\\
m_s(\mu)&=&m_s({\rm 2GeV} )\left[\frac{\alpha_{s}(\mu)}{\alpha_{s}({\rm 2GeV})}\right]^{\frac{12}{33-2n_f}}\, ,\nonumber\\
\alpha_s(\mu)&=&\frac{1}{b_0t}\left[1-\frac{b_1}{b_0^2}\frac{\log t}{t} +\frac{b_1^2(\log^2{t}-\log{t}-1)+b_0b_2}{b_0^4t^2}\right]\, ,
\end{eqnarray}
  where $t=\log \frac{\mu^2}{\Lambda^2}$, $b_0=\frac{33-2n_f}{12\pi}$, $b_1=\frac{153-19n_f}{24\pi^2}$, $b_2=\frac{2857-\frac{5033}{9}n_f+\frac{325}{27}n_f^2}{128\pi^3}$,  $\Lambda=210\,\rm{MeV}$, $292\,\rm{MeV}$  and  $332\,\rm{MeV}$ for the flavors  $n_f=5$, $4$ and $3$, respectively \cite{PDG,Narison-mix}, and evolve all the input parameters to the typical energy scales  $\mu$ with the flavor number $n_f=4$, which satisfy the energy scale formula $\mu=\sqrt{M^2_{X/Y/Z}-(2{\mathbb{M}}_c)^2}$ with the updated value of the effective charmed quark mass ${\mathbb{M}}_c=1.82\,\rm{GeV}$ \cite{WangEPJC-1601,WZG-hidden-charm-PRD}, to extract the masses of the hidden-charm tetraquark states. We tentatively assign the $X(4140)$ and $X(4685)$ to be the ground state and first radial excited state of the hidden-charm tetraquark states respectively, the corresponding pertinent energy scales of the spectral densities at the quark-gluon level  are $\mu=2.0\,\rm{GeV}$ and $3.0\,\rm{GeV}$, respectively.

\begin{table}
\begin{center}
\begin{tabular}{|c|c|c|c|c|c|c|c|}\hline\hline
               &$T^2(\rm{GeV}^2)$   &$\sqrt{s_0}(\rm{GeV})$  &$\mu(\rm{GeV})$  &pole          &$M(\rm{GeV})$  &$\lambda(\rm{GeV}^5)$ \\ \hline

$X(4140)$      &$2.7-3.3$           &$4.7\pm0.1$             &2.0              &$(41-69)\%$   &$4.14\pm0.10$  &$(4.30\pm0.85)\times10^{-2}$   \\ \hline

$X(4685)$      &$2.7-3.3$           &$5.1\pm0.1$             &3.0              &$(69-90)\%$   &$4.70\pm0.12$  &$(1.08\pm0.17)\times10^{-1}$   \\ \hline\hline

\end{tabular}
\end{center}
\caption{ The Borel  windows, continuum threshold parameters, ideal energy scales of the spectral densities, pole contributions,   masses and pole residues for the axialvector   tetraquark states. }\label{mass-tab-X4140-X4685}
\end{table}

 In Ref.\cite{WZG-Di-X4140-EPJC}, we obtain the Borel  window $T^2=2.7-3.3\,\rm{GeV}^2$, continuum threshold parameter $\sqrt{s_0}=4.7\pm0.1\,\rm{GeV}$,  and pole contribution $(41-69)\%$ for the $X(4140)$, then acquire  the mass and pole residue $M_{X}=4.14\pm0.10\,\rm{GeV}$ and $\lambda_X=(4.30\pm0.85)\times10^{-2}\,\rm{GeV}^5$, which are all shown plainly in Table \ref{mass-tab-X4140-X4685}.
 In the present work, we choose the same Borel parameter $T^2=2.7-3.3\,\rm{GeV}^2$ as in Ref.\cite{WZG-Di-X4140-EPJC}, and assume that the energy gap between the first and second radial excited states is about $0.3\sim0.5\,\rm{GeV}$, and take the continuum threshold parameter as $\sqrt{s_0^\prime}=5.1\pm0.1\,\rm{GeV}$, then obtain the pole contribution $(69-90)\%$, it is large enough to extract the mass of the first  radial excited state. Moreover, the convergent behaviors of the operator product expansion are very good, the contributions from the vacuum condensates of dimension 10  in the QCD sum rules for the $X(4140)$ and $X(4685)$ are  $< 1\%$ and $\ll 1\%$, respectively.

\begin{figure}
 \centering
  \includegraphics[totalheight=8cm,width=12cm]{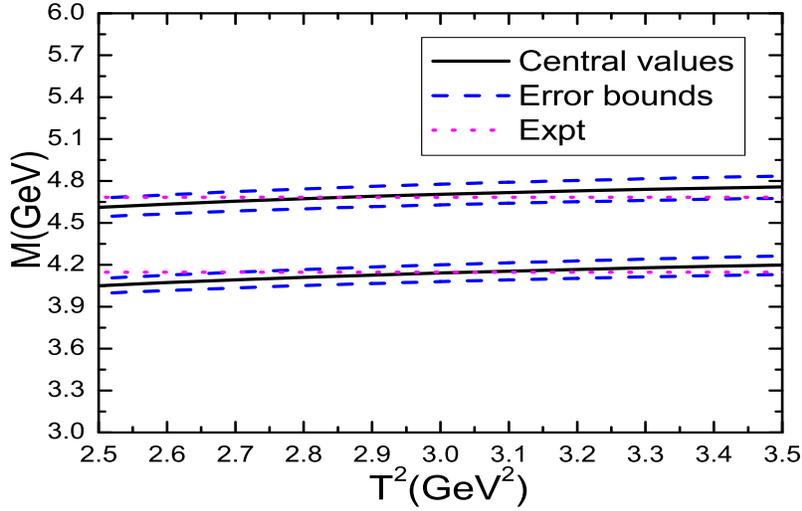}
    \caption{ The masses of the $X$ and $X^\prime$  with variations of the Borel parameter $T^2$,  where  the expt stands for the experimental values of the masses of the $X(4140)$ and $X(4685)$, respectively.   }\label{mass-1S2S-Borel}
\end{figure}

 Finally, we take into account all the uncertainties of the input parameters, and get the mass and pole residue of the first radial excited state $X^\prime$, which are shown plainly in Table \ref{mass-tab-X4140-X4685} and Fig.\ref{mass-1S2S-Borel}. From the Table, we observe that the predicted mass $M_{X^\prime}=4.70\pm0.12\,\rm{GeV}$ is in very good agreement with the experimental value $M_{X(4685)}=4684 \pm 7 {}^{+13}_{-16}\,\rm{MeV}$ from the LHCb collaboration \cite{LHCb-X4685}.

 In Fig.\ref{mass-1S2S-Borel}, we plot the predicted masses $M_{X}$ and $M_{X^\prime}$ with variations of the Borel parameter $T^2$, from the figure, we can see clearly that there appear rather flat platforms both for the ground state and first radial excited state, we are confidential to obtain reliable predictions. In addition, we present the experimental values of the masses of the $X(4140)$ and $X(4685)$, which happen to lie in the center regions  of the predicted values.

 If the masses of the ground state $X(4140)$,  first radial excited state $X(4685)$, second radial excited state $X^{\prime\prime}$, etc satisfy the  Regge trajectory,
 \begin{eqnarray}
 M_n^2&=&\alpha (n-1)+\alpha_0\, ,
 \end{eqnarray}
 where the $\alpha$ and $\alpha_0$ are some constants to be fitted experimentally, the $n$ is the radial quantum number. We take the masses of the ground state and first radial excited state, $M_{X(4140)}=4118\,\rm{MeV}$ and $M_{X(4685)}=4684\,\rm{MeV}$ \cite{LHCb-X4685}, as input parameters to fit the parameters $\alpha$ and $\alpha_0$, and obtain the mass of the second radial excited state, $M_{X^{\prime\prime}}=5.19\pm0.10\,\rm{GeV}$, which
 is consistent with the continuum threshold parameter $\sqrt{s_0^\prime}=5.1\pm0.1\,\rm{GeV}$, the contamination from the second radial excited state is avoided, here we add an  uncertainty $\delta=\pm 0.1\,\rm{GeV}$ to the mass $M_{X^{\prime\prime}}$ according to Table \ref{mass-tab-X4140-X4685}. Now we reach the conclusion tentatively that the calculations are self-consistent.

 The values $M_{X^{\prime\prime}}=5.09\,\rm{GeV}$, $5.19\,\rm{GeV}$ and $5.29\,\rm{GeV}$ correspond to the continuum threshold parameters $\sqrt{s_0^\prime}=5.0\,\rm{GeV}$, $5.1\,\rm{GeV}$ and $5.2\,\rm{GeV}$, respectively, and have the relation $M_{X^{\prime\prime}}>\sqrt{s_0^\prime}$, the contamination from the second radial excited state can be neglected. At the beginning, we assume that the energy gap between the first and second radial excited states is about $0.3\sim0.5\,\rm{GeV}$, and tentatively take the continuum threshold parameter as $\sqrt{s_0^\prime}=5.1\pm0.1\,\rm{GeV}$ to obtain the mass of the $X^\prime$, then resort to the Regge trajectory to check whether or not such a choice is self-consistent.  Fortunately, such a choice happens to be satisfactory. On the other hand, if it is not self-consistent, we can choose another value of the $\sqrt{s_0^\prime}$, then repeat the same routine to obtain self-consistent $M_{X^{\prime}}$, $M_{X^{\prime\prime}}$ and $\sqrt{s_0^\prime}$ via trial and error. In all the calculations, we should obtain flat  Borel platforms to suppress the dependence on the Borel parameters.

In Refs.\cite{WZG-Landau,WZG-ESF-EPJC-2874,X3915-X4500-EPJC-WZG,WZG-Di-X4140-EPJC}, we investigate  the hidden-charm tetraquark (molecular) states via the QCD sum rules using the energy scale formula  $\mu=\sqrt{M^2_{X/Y/Z}-(2{\mathbb{M}}_c)^2}$ to choose the pertinent energy scales of the spectral densities at the quark and gluon level, which can enhance the pole contributions remarkably and improve the convergent behaviors of the operator product expansion remarkably. It is a unique  feature of our works.
The predictions $M_{D_s^*\bar{D}_{s1}-D_{s1}\bar{D}_s^*}=4.67\pm0.08\,\rm{GeV}$ \cite{WZG-Landau},
$M_{[cs]_P[\bar{c}\bar{s}]_A+[cs]_A[\bar{c}\bar{s}]_P}=4.63^{+0.11}_{-0.08}\, \rm{GeV}$ \cite{WZG-ESF-EPJC-2874},
$M_{[cs]_A[\bar{c}\bar{s}]_A,\rm 1S}=3.92^{+0.19}_{-0.18}\,\rm{GeV}$,
$M_{[cs]_A[\bar{c}\bar{s}]_A,\rm 2S}=4.50^{+0.08}_{-0.09}\,\rm{GeV}$ \cite{X3915-X4500-EPJC-WZG},
$M_{[sc]_{\widetilde{V}}[\bar{s}\bar{c}]_V-[sc]_V[\bar{s}\bar{c}]_{\widetilde{V}},\rm 1S}=4.14\pm0.10\,\rm{GeV}$ \cite{WZG-Di-X4140-EPJC} and
$M_{[sc]_{\widetilde{V}}[\bar{s}\bar{c}]_V-[sc]_V[\bar{s}\bar{c}]_{\widetilde{V}},\rm 2S}=4.70\pm0.12\,\rm{GeV}$
based on the QCD sum rules support assigning the $X(4630)$ as the $D_s^*\bar{D}_{s1}-D_{s1}\bar{D}_s^*$ tetraquark molecular state \cite{WZG-Landau} or   $[cs]_P[\bar{c}\bar{s}]_A+[cs]_A[\bar{c}\bar{s}]_P$ tetraquark state with the quantum numbers $J^{PC}=1^{-+}$ \cite{WZG-ESF-EPJC-2874}, assigning the $X(3915)$ and $X(4500)$ as the 1S and 2S $[cs]_A[\bar{c}\bar{s}]_A$ tetraquark states respectively with the  quantum numbers $J^{PC}=0^{++}$, and assigning the $X(4140)$ and $X(4685)$
 as the 1S and 2S $[sc]_{\widetilde{V}}[\bar{s}\bar{c}]_V-[sc]_V[\bar{s}\bar{c}]_{\widetilde{V}}$ tetraquark states respectively with the  quantum numbers $J^{PC}=1^{++}$ \cite{WZG-Di-X4140-EPJC}. The predictions with the possible assignments are given  plainly in Table \ref{1S2S-assignment}. We should bear in mind that other assignments of the $X(4500)$ and $X(4700)$, such as the D-wave $cs\bar{c}\bar{s}$ tetraquark states with the $J^P=0^+$ are also possible \cite{Chen-Chen-Zhu-4500}, more theoretical and experimental works are still needed to obtain definite conclusion.

\begin{table}
\begin{center}
\begin{tabular}{|c|c|c|c|c|c|c|c|}\hline\hline
                                                       &$J^{PC}$   & 1S [mass\,(GeV)]                   & 2S [mass\,(GeV)]  & References     \\ \hline

 $D_s^*\bar{D}_{s1}-D_{s1}\bar{D}_s^*$                 &$1^{-+}$   & $X(4630)$ [$4.67\pm0.08$]          &                   & \cite{WZG-Landau} \\ \hline

 $[cs]_P[\bar{c}\bar{s}]_A+[cs]_A[\bar{c}\bar{s}]_P$   &$1^{-+}$   & $X(4630)$ [$4.63^{+0.11}_{-0.08}$] &                   & \cite{WZG-ESF-EPJC-2874} \\ \hline

 $[cs]_A[\bar{c}\bar{s}]_A$              & $0^{++}$  & $X(3915)$ [$3.92^{+0.19}_{-0.18}$]  & $X(4500)$ [$4.50^{+0.08}_{-0.09}$] &\cite{X3915-X4500-EPJC-WZG}   \\ \hline

 $[sc]_{\widetilde{V}}[\bar{s}\bar{c}]_V-[sc]_V[\bar{s}\bar{c}]_{\widetilde{V}}$  &$1^{++}$   & $X(4140)$ [$4.14\pm0.10$]  & $X(4685)$ [$4.70\pm0.12$] &\cite{WZG-Di-X4140-EPJC}${}^*$   \\ \hline \hline
\end{tabular}
\end{center}
\caption{ The possible assignments of the LHCb's $X$ states based on the predictions from the QCD sum rules, where the superscript $*$ denotes the mass of the $X(4140)$  taken from Ref.\cite{WZG-Di-X4140-EPJC}. }\label{1S2S-assignment}
\end{table}

Now we explore the outcome in the case of saturating the hadron side of the QCD sum rules with  the two-meson scattering states.
At the hadron side of the QCD sum rules in Eqs.\eqref{TM-QCDSR-1}-\eqref{TM-QCDSR-2}, we choose the  parameters
$m_{J/\psi}=3.0969\,\rm{GeV}$,
$m_{\eta_c}=2.9839\,\rm{GeV}$,
$m_{h_c}=3.52538\,\rm{GeV}$,
$m_{\chi_{c0}}=3.41471\,\rm{GeV}$,
$m_{\chi_{c1}}=3.51067\,\rm{GeV}$,
$m_{\phi}=1.019461\,\rm{GeV}$,
$m_{h_1}=1.416\,\rm{GeV}$,
$m_{f_1}=1.4263\,\rm{GeV}$,
$m_{f_0}=1.506\,\rm{GeV}$,
$m_{\eta}=0.547862\,\rm{GeV}$,
$m_{D_s}=1.969\,\rm{GeV}$,
$m_{D_s^*}=2.1122\,\rm{GeV}$,
$m_{D_{s0}}=2.318\,\rm{GeV}$,
$m_{D_{s1}}=2.4596\,\rm{GeV}$,
$m_{\psi^\prime}=3.6861\,\rm{GeV}$,
$m_{\psi^{\prime\prime}}=4.0396\,\rm{GeV}$ from the Particle Data Group \cite{PDG};
$m_{h^\prime_c}=3.9560\,\rm{GeV}$ from  the Godfrey-Isgur model  \cite{Godfrey-hc};
$f_{J/\psi}=0.418 \,\rm{GeV}$,
$f_{\eta_c}=0.387 \,\rm{GeV}$,
$f_{J/\psi}^T=0.410 \,\rm{GeV}$,
$f_{h_c}=0.235 \,\rm{GeV}$
 from the Lattice QCD \cite{Becirevic};
 $f_{\chi_{c1}}=0.338 \,\rm{GeV}$, $f_{\chi_{c0}}=0.359 \,\rm{GeV}$ \cite{VANovikov-PRT},
 $f_{\phi}=0.231 \,\rm{GeV}$,
 $f_{\phi}^T=0.200 \,\rm{GeV}$ \cite{PBall2004,Wang-Y4274},
 $f_\eta=1.34 f_\pi$ \cite{Feldmann},
   $f_{h_1}=0.183 \,\rm{GeV}$, $f_{f_1}=0.211 \,\rm{GeV}$ \cite{KCYang},
   $f_{f_0}=0.490 \,\rm{GeV}$ \cite{HYCheng},
 $f_{D_s}=0.240 \,\rm{GeV}$,
$f_{D_s^*}=0.308 \,\rm{GeV}$,
$f_{D_{s0}}=0.333 \,{\rm{GeV}} \frac{m_c}{m_{D_{s0}}}$,
$f_{D_{s1}}=0.345 \,\rm{GeV}$
 \cite{Wang-DecayConst} from the QCD sum rules;
 $f_{\pi}=0.130 \,\rm{GeV}$, $f_{\psi^\prime}=0.295 \,\rm{GeV}$, $f_{\psi^{\prime\prime}}=0.187 \,\rm{GeV}$ extracted from the experimental data \cite{PDG};
 $f_{D_s^*}^T=f_{D_s^*}$, $f_{\psi^\prime}^T=f_{\psi^\prime}$, $f^T_{\psi^{\prime\prime}}=f_{\psi^{\prime\prime}}$,
  $f_{h_c^\prime}=f_{h_c}\frac{f_{\psi^\prime}}{f_{J/\psi}}=0.166\,\rm{GeV}$ estimated in the present work (also in Ref.\cite{WZG-IJMPA-nonlocal}).

\begin{figure}
 \centering
  \includegraphics[totalheight=8cm,width=10cm]{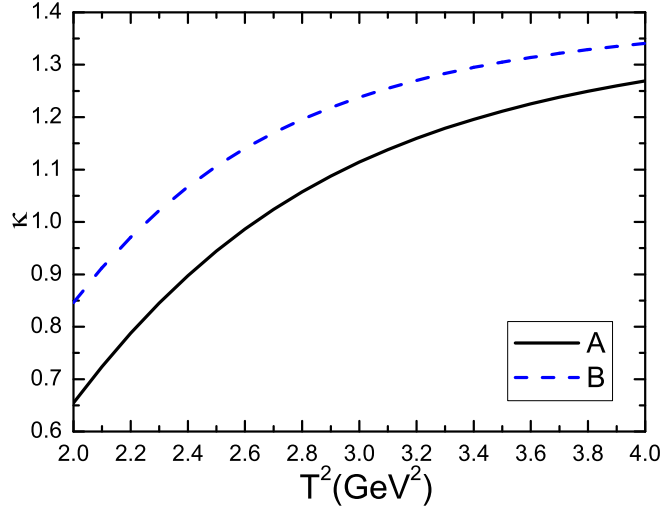}
    \caption{ The values of the $\kappa$   with variations of the Borel parameter $T^2$,  where  the $A$ and $B$ come from the QCD sum rules in Eq.\eqref{TM-QCDSR-1} and Eq.\eqref{TM-QCDSR-2}, respectively.   }\label{kappa-Borel}
\end{figure}

In Fig.\ref{kappa-Borel}, we  plot the values of the $\kappa$    with variations of the  Borel parameter $T^2$ for the central values of the input parameters.
From Fig.\ref{kappa-Borel}, we can see that the values of the $\kappa$ increase monotonically and
 quickly  with the increase of the Borel parameter $T^2$, no platform appears, which indicates  that the QCD sum rules in Eqs.\eqref{TM-QCDSR-1}-\eqref{TM-QCDSR-2} are not satisfactory, the two-meson scattering states alone cannot saturate the QCD sum rules at the hadron side.

In Ref.\cite{WZG-IJMPA-nonlocal}, we investigate the $Z_c(3900)$ with the QCD sum rules in details by including all the two-meson scattering state contributions   and nonlocal effects between the diquark and antidiquark constituents. We observe that the two-meson scattering states alone cannot saturate  the QCD sum rules at the hadron side, just like in the present case,  the contribution of the $Z_c(3900)$ (or pole term) plays an un-substitutable role, we can saturate the QCD sum rules with or without the two-meson  scattering state contributions. We expect the conclusion is also applicable in the present case.

Now we explore the two-meson scattering state contributions besides the tetraquark states $X_c$ and $X_c^\prime$, and take account of all the contributions,
\begin{eqnarray}\label{Self-Energy}
\Pi_{\mu\nu}(p) &=&-\frac{\widehat{\lambda}_{X}^{2}}{ p^2-\widehat{M}_{X}^2+\Sigma_{J/\psi\phi}(p^2)+\cdots}g_{\mu\nu}-\frac{\widehat{\lambda}_{X^\prime}^{2}}{ p^2-\widehat{M}_{X^\prime}^2+\Sigma_{J/\psi\phi}(p^2)+\cdots}g_{\mu\nu}+\cdots \, ,
\end{eqnarray}
we choose the bare masses and pole residues $\widehat{M}_{X}$, $\widehat{M}_{X^\prime}$, $\widehat{\lambda}_{X}$ and $\widehat{\lambda}_{X^\prime}$  to absorb the divergent terms  in the self-energies  $\Sigma_{J/\psi \phi}(p^2)$, $\cdots$. The renormalized self-energies  satisfy  the relations  $p^2-M_{X,R}^2+\overline{\Sigma}_{J/\psi\phi}(p^2)+\cdots=0$ and $p^2-M_{X^\prime,R}^2+\overline{\Sigma}_{J/\psi\phi}(p^2)+\cdots=0$, where the subscripts $R$ represent  the $\overline{MS}$ masses, the overlines above the
self-energies represent  that the divergent terms have been subtracted. The tetraquark states $X_c$ and $X_c^\prime$ have finite widths and are unstable particles, the relations should be modified,
$p^2-M_{X,R}^2+{\rm Re}\overline{\Sigma}_{J/\psi \phi}(p^2)+\cdots=0$, $p^2-M_{X^\prime,R}^2+{\rm Re}\overline{\Sigma}_{J/\psi \phi}(p^2)+\cdots=0$,
${\rm Im}\overline{\Sigma}_{J/\psi\phi}(p^2)+\cdots=\sqrt{p^2}\Gamma_{X}(p^2)$, and ${\rm Im}\overline{\Sigma}_{J/\psi\phi}(p^2)+\cdots=\sqrt{p^2}\Gamma_{X^\prime}(p^2)$. The renormalized self-energies  contribute  a finite imaginary part to modify the dispersion relation,
\begin{eqnarray}
\Pi_{\mu\nu}(p) &=&-\frac{\lambda_{X}^{2}}{ p^2-M_{X}^2+i\sqrt{p^2}\Gamma_X(p^2)}g_{\mu\nu}-\frac{\lambda_{X^\prime}^{2}}{ p^2-M_{X^\prime}^2+i\sqrt{p^2}\Gamma_{X^\prime}(p^2)}g_{\mu\nu}+\cdots \, ,
 \end{eqnarray}
where $M_{X^{(\prime)}}^2=M_{X^{(\prime)},R}^2+\overline{\Sigma}_{J/\psi\phi}(M_{X^{(\prime)}}^2)$.

We can take  account of the finite width effects by the  simple replacements of the hadronic spectral densities,
\begin{eqnarray}
\lambda^2_{X^{(\prime)}}\delta \left(s-M^2_{X^{(\prime)}} \right) &\to& \lambda^2_{X^{(\prime)}}\frac{1}{\pi}\frac{M_{X^{(\prime)}}\Gamma_{X^{(\prime)}}(s)}{(s-M_{X^{(\prime)}}^2)^2+M_{X^{(\prime)}}^2\Gamma_{X^{(\prime)}}^2(s)}\, ,
\end{eqnarray}
where
\begin{eqnarray}
\Gamma_{X^{(\prime)}}(s)&=&\Gamma_{X^{(\prime)}} \frac{M_{X^{(\prime)}}}{\sqrt{s}}\sqrt{\frac{s-(m_{J/\psi}+m_{\phi})^2}{M^2_{X^{(\prime)}}-(m_{J/\psi}+m_{\phi})^2}}
\, ,
\end{eqnarray}
the $\Gamma_{X}$ and $\Gamma_{X^\prime}$ are the physical decay widths.
Then the hadron  sides of  the QCD sum rules  undergo the following changes,
\begin{eqnarray}
\lambda^2_{X^{(\prime)}}\exp \left(-\frac{M^2_{X^{(\prime)}}}{T^2} \right) &\to& \lambda^2_{X^{(\prime)}}\int_{(m_{J/\psi}+m_{\phi})^2}^{s_0^{(\prime)}}ds\frac{1}{\pi}
\frac{M_{X^{(\prime)}}\Gamma_{X^{(\prime)}}(s)}{(s-M_{X^{(\prime)}}^2)^2+M_{X^{(\prime)}}^2\Gamma_{X^{(\prime)}}^2(s)}\exp \left(-\frac{s}{T^2} \right)\, , \nonumber\\
&=&\left(0.81\sim0.82\lambda_{X}\right)^2  \,\, (\left(0.97\sim0.98\lambda_{X^\prime}\right)^2  \exp \left(-\frac{M^2_{X^{(\prime)}}}{T^2} \right)\, , \\
\lambda^2_{X^{(\prime)}}M^2_{X^{(\prime)}}\exp \left(-\frac{M^2_{X^{(\prime)}}}{T^2} \right) &\to& \lambda^2_{X^{(\prime)}}\int_{(m_{J/\psi}+m_{\phi})^2}^{s_0^{(\prime)}}ds\,s\,\frac{1}{\pi}\frac{M_{X^{(\prime)}}\Gamma_{X^{(\prime)}}(s)}{(s-M_{X^{(\prime)}}^2)^2
+M_{X^{(\prime)}}^2\Gamma_{X^{(\prime)}}^2(s)}\exp \left(-\frac{s}{T^2} \right)\, , \nonumber\\
&=&\left(0.82\sim0.83\lambda_{X}\right)^2  \,\, (\left(0.96\sim0.97\lambda_{X^\prime}\right)^2\,M^2_{X^{(\prime)}}\exp \left(-\frac{M^2_{X^{(\prime)}}}{T^2} \right)\, .
\end{eqnarray}
We can absorb the numerical factors  $0.81\sim0.82$, $0.82\sim0.83$, $0.97\sim0.98$ and $0.96\sim0.97$ into the pole residues  safely, the two-meson scattering states   cannot  affect  the masses $M_{X}$ and $M_{X^\prime}$ significantly \cite{WZG-IJMPA-4200}. Again, we obtain the conclusion, the pole terms or tetraquark states play an un-substitutable role, we can saturate the QCD sum rules with or without the two-particle scattering state contributions, the two-particle scattering states can only modify the pole residues \cite{WZG-IJMPA-nonlocal}.

In the present work, we choose the local four-quark current $J_\mu(x)$, while the traditional mesons are spatial extended objects and have average  spatial sizes $\sqrt{\langle r^2\rangle} \neq 0$, for example,
 $\sqrt{\langle r^2\rangle}=0.41\,\rm{fm}$ ($0.42\,\rm{fm}$)  for the $J/\psi$ \cite{KTChao-0903}  (\cite{Buchmuller-1981}), $\sqrt{\langle r^2\rangle}=0.63\,\rm{fm}$ for the $\phi(1020)$ \cite{ZhongXH-phi}. On the other hand,  the diquark-antidiquark type tetraquark states have the  average  spatial sizes $\langle r\rangle=0.5\sim 0.7\,\rm{fm}$ \cite{RFLebed-tetra-radius}. The $J/\psi$, $\phi(1020)$, $X(4140)$ and $X(4685)$ have average  spatial sizes of the same order, the couplings to the continuum states $J/\psi \phi$ et al can be neglected, as the overlappings  of the wave-functions are small enough.

\section{Conclusion}

At the first step, we take into account our previous calculations based on the QCD sum rules and make possible assignments of the LHCb's new particles $X(4630)$ and $X(4500)$. We tentatively   assign the $X(4630)$ to be the $D_s^*\bar{D}_{s1}-D_{s1}\bar{D}_s^*$ tetraquark molecular state  or   $[cs]_P[\bar{c}\bar{s}]_A+[cs]_A[\bar{c}\bar{s}]_P$ tetraquark state with the quantum numbers $J^{PC}=1^{-+}$, and assign  the $X(3915)$ and $X(4500)$ to be the 1S and 2S $[cs]_A[\bar{c}\bar{s}]_A$ tetraquark states respectively with the  quantum numbers $J^{PC}=0^{++}$ according to the predicted masses.

Then we extend our  previous works to explore  the $X(4685)$ as the first radial excited state of the $X(4140)$ with the QCD sum rules, and obtain the value of the mass $M_{X(4685)}=4.70\pm0.12\,\rm{GeV}$, which is in very good agreement with the experimental value $M_{X(4685)}=4684 \pm 7 {}^{+13}_{-16}\,\rm{MeV}$ from the LHCb collaboration, and supports   assigning the $X(4140)$ and $X(4685)$
 as the 1S and 2S $[sc]_{\widetilde{V}}[\bar{s}\bar{c}]_V-[sc]_V[\bar{s}\bar{c}]_{\widetilde{V}}$ tetraquark states respectively with the  quantum numbers $J^{PC}=1^{++}$. Furthermore,   we investigate  the two-meson scattering state contributions  in details, and observe that the two-meson scattering state contributions alone cannot saturate the QCD sum rules at the hadron side,  the contributions of the tetraquark states (or pole terms) play an un-substitutable role, we can saturate the QCD sum rules with or without the two-meson scattering state contributions,  the two-meson scattering state contributions can only modify the pole residues, the predictions of the tetraquark masses are robust.

\section*{Acknowledgements}
This  work is supported by National Natural Science Foundation, Grant Number  11775079.


\begin{thebibliography}{99}

\bibitem{CDF0903}  T. Aaltonen et al,   Phys. Rev. Lett. {\bf 102} (2009) 242002.


\bibitem{CDF1101} T. Aaltonen et al,  Mod. Phys. Lett. {\bf A32} (2017) 1750139.

\bibitem{CMS1309} S. Chatrchyan et al, Phys. Lett. {\bf B734} (2014) 261.

\bibitem{D0-1309}  V. M. Abazov et al,  Phys. Rev. {\bf D89} (2014) 012004.

\bibitem{D0-1508}  V. M. Abazov et al, Phys. Rev. Lett. {\bf 115} (2015) 232001.

\bibitem{LHCb-16061}  R. Aaij et al,  Phys. Rev. Lett. {\bf 118} (2017) 022003.

\bibitem{LHCb-16062}  R. Aaij et al,  Phys. Rev. {\bf D95} (2017) 012002.

\bibitem{LHCb-X4685} R. Aaij et al, arXiv:2103.01803 [hep-ex].


\bibitem{WZG-Landau} Z. G. Wang, Phys. Rev. {\bf D101} (2020)  074011.


\bibitem{FKGuo-Progr} X. K. Dong, F. K. Guo and B. S. Zou, Progr. Phys. {\bf 41} (2021) 65.

\bibitem{DYChen-EPJC-2020} J. He, Y. Liu, J. T. Zhu and D. Y. Chen, Eur. Phys. J. {\bf C80} (2020)  246.


\bibitem{LiuX-X4630} X. D. Yang, F. L. Wang, Z. W. Liu and X. Liu, arXiv:2103.03127.


\bibitem{WZG-ESF-EPJC-2874} Z. G. Wang, Eur. Phys. J. {\bf C74} (2014)  2874.


\bibitem{X3915-X4500-EPJC-WZG} Z. G. Wang, Eur. Phys. J. {\bf C77} (2017)  78.

\bibitem{X3915-X4500-EPJA-WZG}   Z. G. Wang,  Eur. Phys. J. {\bf A53} (2017) 19.

\bibitem{Chen-Chen-Zhu-4500} H. X. Chen, E. L. Cui, W. Chen, X. Liu and S. L. Zhu, Eur. Phys. J. {\bf C77} (2017)  160.


\bibitem{X4140-tetraquark-Lebed} R. F. Lebed and A. D. Polosa,  Phys. Rev. {\bf D93} (2016) 094024.



\bibitem{Maiani-Z4430-1405} L. Maiani, F. Piccinini, A. D. Polosa and V. Riquer, Phys. Rev. {\bf D89} (2014) 114010.


\bibitem{Nielsen-1401} M. Nielsen and F. S. Navarra,  Mod. Phys. Lett. {\bf  A29} (2014) 1430005.

\bibitem{WangZG-Z4430-CTP} Z. G. Wang,  Commun. Theor. Phys. {\bf 63} (2015)  325.


\bibitem{ChenHX-Z4600-A} H. X. Chen and W. Chen,  Phys. Rev. {\bf D99} (2019)  074022.


\bibitem{WangZG-axial-Z4600} Z. G. Wang, Chin. Phys. {\bf C44} (2020) 063105.

\bibitem{WZG-Di-X4140-EPJC} Z. G. Wang and Z. Y. Di, Eur. Phys. J. {\bf C79} (2019) 72.

\bibitem{Wilczek-diquark} A. Selem and F. Wilczek,  hep-ph/0602128.

\bibitem{Polosa-diquark} L. Maiani, A. D. Polosa and V. Riquer, Phys. Lett. {\bf B778} (2018) 247.

\bibitem{Brodsky-PRL} S. J. Brodsky, D. S. Hwang and R. F. Lebed, Phys. Rev. Lett. {\bf 113} (2014)  112001.

\bibitem{WZG-IJMPA-nonlocal} Z. G. Wang, Int. J. Mod. Phys. {\bf A35} (2020)  2050138.

\bibitem{WZG-HT-PRD-2014} Z. G. Wang and T. Huang, Phys. Rev. {\bf D89} (2014) 054019.

\bibitem{WZG-HT-EPJC-2891} Z. G. Wang and T. Huang, Eur. Phys. J. {\bf C74} (2014)  2891.

\bibitem{WZG-EPJC-2963} Z. G. Wang, Eur. Phys. J. {\bf C74} (2014) 2963.


\bibitem{Novikov--shifman} V. A. Novikov, M. A. Shifman, A. I. Vainshtein, M. B. Voloshin and V. I. Zakharov, Nucl. Phys. {\bf B237} (1984) 525.

\bibitem{Narison-DvDv-D-8} R. Albuquerque, S. Narison, F. Fanomezana, A. Rabemananjara, D. Rabetiarivony and G. Randriamanatrika,
Int. J. Mod. Phys. {\bf A31} (2016) 1650196.


\bibitem{SVZ79} M. A. Shifman, A. I. Vainshtein and V. I. Zakharov, Nucl. Phys. {\bf B147} (1979) 385; Nucl. Phys. {\bf B147} (1979) 448.

\bibitem{Review-rho-kappa} D.  B. Leinweber, Annals Phys. {\bf 254} (1997) 328; and references  therein.

\bibitem{Narison-rho} S. Narison,  Phys. Lett. {\bf B673} (2009) 30.


\bibitem{WZG-DvDvDv} Z. G. Wang, Commun. Theor. Phys. {\bf 73} (2021)  065201.

\bibitem{Gubler-SB} P. Gubler and D. Satow, Prog. Part. Nucl. Phys. {\bf 106} (2019) 1.


\bibitem{Baxi-G} M. S. Maior de Sousa and R. Rodrigues da Silva, Braz. J. Phys. {\bf 46} (2016) 730.

\bibitem{Reinders85} L. J. Reinders, H. Rubinstein and S. Yazaki, Phys. Rept. {\bf 127} (1985) 1.


\bibitem{ColangeloReview} P. Colangelo and A. Khodjamirian, hep-ph/0010175.

\bibitem{PDG}  P. A. Zyla et al,  Prog. Theor. Exp. Phys. {\bf 2020} (2020) 083C01.


\bibitem{Narison-mix} S. Narison and R. Tarrach, Phys. Lett. {\bf 125 B} (1983) 217.

\bibitem{WangEPJC-1601} Z. G. Wang, Eur. Phys. J. {\bf C76} (2016)  387.


\bibitem{WZG-hidden-charm-PRD} Z. G. Wang, Phys. Rev. {\bf D102} (2020) 014018.



\bibitem{Godfrey-hc} T. Barnes, S. Godfrey and E. S. Swanson, Phys. Rev. {\bf D72} (2005) 054026.



\bibitem{Becirevic} D. Becirevic, G. Duplancic, B. Klajn, B. Melic and F. Sanfilippo,  Nucl. Phys. {\bf B883} (2014) 306.

\bibitem{VANovikov-PRT} V. A. Novikov, L. B. Okun, M. A. Shifman, A. I. Vainshtein, M. B. Voloshin and V. I. Zakharov, Phys. Rept. {\bf 41} (1978) 1.


\bibitem{PBall2004} P. Ball and R. Zwicky, Phys. Rev. {\bf D71} (2005) 014029.

\bibitem{Wang-Y4274}  Z. G. Wang, Eur. Phys. J. {\bf C77} (2017)  174.


\bibitem{Feldmann} T. Feldmann, P. Kroll and B. Stech, Phys. Rev. {\bf D58} (1998) 114006.


\bibitem{KCYang} K. C. Yang, Nucl. Phys. {\bf B776} (2007) 187.

\bibitem{HYCheng} H. Y. Cheng, C. K. Chua and K. C. Yang,  Phys. Rev. {\bf D73} (2006 ) 014017.

\bibitem{Wang-DecayConst}  Z. G. Wang, Eur. Phys. J. {\bf C75} (2015) 427.


\bibitem{WZG-IJMPA-4200} Z. G. Wang, Int. J. Mod. Phys. {\bf A30} (2015)  1550168.

\bibitem{KTChao-0903}  B. Q. Li and K. T. Chao, Phys. Rev. {\bf D79} (2009) 094004.

\bibitem{Buchmuller-1981} W. Buchmuller and S. H. H. Tye, Phys. Rev. {\bf D24} (1981) 132.

\bibitem{ZhongXH-phi} X. H. Zhong, private communication.

\bibitem{RFLebed-tetra-radius} J. F. Giron, R. F. Lebed and C. T. Peterson, JHEP {\bf 2001} (2020) 124.


\end{thebibliography}
\end{document}